\documentclass{ws-mpla}

\def\ga{\gamma}
\def\de{\delta}

\def\th{\theta}

\def\ka{\kappa}
\def\la{\lambda}

\def\si{\sigma}

\def\ph{\phi}

\def\ch{\chi}
\def\ps{\psi}
\def\om{\omega}
\def\Ga{\Gamma}
\def\De{\Delta}

\def\La{\Lambda}

\def\nue{\nu_e}
\def\numu{\nu_\mu}
\def\nubar{\bar\nu}
\def\nuebar{\bar\nu_e}
\def\numubar{\bar\nu_\mu}

\def\fr#1#2{\frac{#1}{#2}}

\def\lsim{\mathrel{\rlap{\lower4pt\hbox{\hskip1pt$\sim$}}
    \raise1pt\hbox{$<$}}}
\def\gsim{\mathrel{\rlap{\lower4pt\hbox{\hskip1pt$\sim$}}
    \raise1pt\hbox{$>$}}}

\newcommand{\beq}{\begin{eqnarray}}
\newcommand{\eeq}{\end{eqnarray}}
\def\to{\rightarrow}

\def\no{\nonumber}

\def\aL{(a_L)}
\def\cL{(c_L)}

\def\As#1{({\cal A}_s)_{#1}}
\def\Ac#1{({\cal A}_c)_{#1}}
\def\Bs#1{({\cal B}_s)_{#1}}
\def\Bc#1{({\cal B}_c)_{#1}}
\def\C#1{({\cal C})_{#1}}
\def\Asa#1{({\cal A}_s^{(0)})_{#1}}
\def\Asc#1{({\cal A}_s^{(1)})_{#1}}
\def\Aca#1{({\cal A}_c^{(0)})_{#1}}
\def\Acc#1{({\cal A}_c^{(1)})_{#1}}
\def\Bsc#1{({\cal B}_s^{(1)})_{#1}}
\def\Bcc#1{({\cal B}_c^{(1)})_{#1}}

\def\Ca#1{({\cal C}^{(0)})_{#1}}
\def\Cc#1{({\cal C}^{(1)})_{#1}}
\def\nh^#1{{\hat N}^{#1}}
\def\indxn{{e\mu}}

\def\mF_#1{{\cal F}_{#1}} 

\def\gaf{{\gamma}_5}
\def\gam{{\gamma}_{\mu}}

\def\ganu{{\gamma}^{\nu}}

\def\MBosc1POT{5.58\times 10^{20}}

\def\tsid{86164.1}
\def\tgmt{86400.0}

\def\tdch{53.9}\def\tddf{49}\def\tdpb{29.3}

\def\MBch{48.2}
\def\MBth{89.8}
\def\MBph{180.0}

\def\Dcsqnu{26.9}
\def\Dcsqnubar{3.0}

\def\np{6.46\times 10^{20}}
\def\nln{544}\def\nlb{409.8}\def\nlt{23.3}\def\nly{38.3}

\def\ap{5.66\times 10^{20}}

\def\atn{241}\def\atb{200.7}\def\att{15.5}\def\aty{14.3}

\begin{document}

\markboth{T. Katori}
{Tests of Lorentz and CPT violation with MiniBooNE neutrino oscillation excesses}

\catchline{}{}{}{}{}

\title{TESTS OF LORENTZ AND CPT VIOLATION WITH MINIBOONE NEUTRINO OSCILLATION EXCESSES}
 
\author{TEPPEI KATORI}
\address{Laboratory for Nuclear Science, Massachusetts Institute of Technology,\\
Cambridge, MA 02139, U.S.A.\\
katori@mit.edu
}

\maketitle

\pub{Received (Day Month Year)}{Revised (Day Month Year)}

\begin{abstract}

Violation of Lorentz invariance and CPT symmetry is a 
predicted phenomenon of Planck-scale physics. 
Various types of data are analyzed to search for Lorentz violation under 
the Standard-Model Extension (SME) framework, 
including neutrino oscillation data.
MiniBooNE is a short-baseline neutrino oscillation experiment at Fermilab. 
The measured excesses from MiniBooNE cannot be reconciled within 
the neutrino Standard Model ($\nu$SM); 
thus it might be a signal of new physics, such as Lorentz violation. 
We have analyzed the sidereal time dependence of MiniBooNE data 
for signals of the possible breakdown of Lorentz invariance in neutrinos.  
In this brief review, we introduce Lorentz violation, 
the neutrino sector of the SME, 
and the analysis of short-baseline neutrino oscillation experiments.  
We then present the results of 
the search for Lorentz violation in MiniBooNE data. 
This review is based on the published result~\cite{MB_LV}.

\end{abstract}

\ccode{PACS: 11.30.Cp, 14.60.Pq, 14.60.St}

\section{Spontaneous Lorentz symmetry breaking (SLSB)}	

Every fundamental symmetry needs to be tested, including Lorentz symmetry. 
The breakdown of Lorentz invariance naturally 
arises in different scenarios of physics at the Planck scale. 
For this reason the expected scale of Lorentz-violating phenomena 
is more than the Planck mass $m_P\simeq 10^{19}$~GeV, 
or in other words, 
Lorentz violation is expected to be suppressed 
until at least $\simeq 10^{-19}$~GeV in our energy scale. 
Lorentz symmetry is a fundamental symmetry both 
in quantum field theory and general relativity; 
the consequence of its violation would be enormous, 
and it seems it is impossible to establish a self-consistent 
theory with Lorentz violation. 
However, Lorentz violation can be incorporated into existing theories 
by spontaneous breaking. In this way, Lorentz-violating terms in 
the Lagrangian do not conflict with the Standard Model (SM).

There are a number of models for 
spontaneous Lorentz symmetry breaking (SLSB),\cite{SLSB} 
although the basic idea is the same for all. 
In the SM, the mass terms arise from the spontaneous symmetry breaking (SSB) 
triggered by a Higgs field with a nonzero vacuum expectation value. 
In this way, mass terms are dynamically generated and 
they do not exist before SSB. 
Consequently, the SM does not have to violate gauge symmetry where 
mass terms would necessarily do so.

\begin{figure}[tbp]
\centerline{\psfig{file=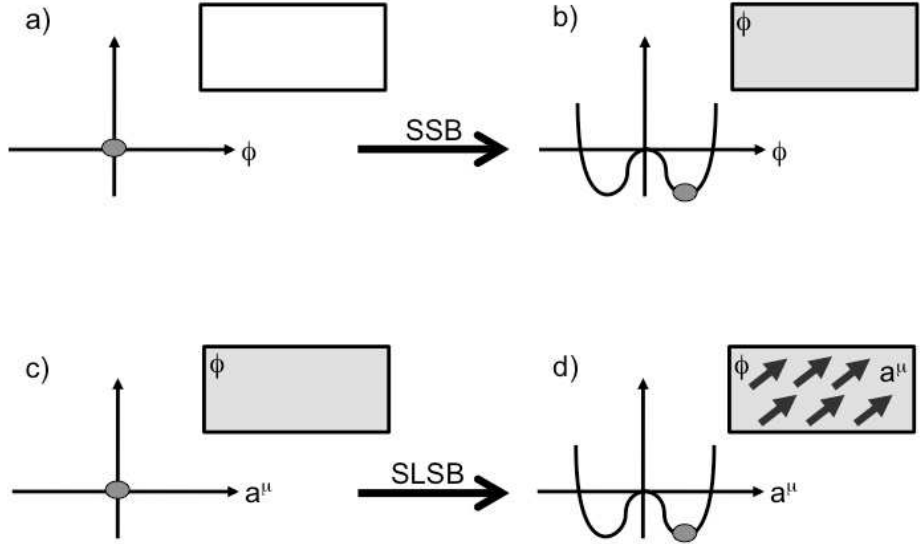,width=5.0in}}
\vspace*{8pt}
\caption{An illustration of spontaneous symmetry breaking (SSB).}
\label{fig:SLSB}
\end{figure}

Figure~\ref{fig:SLSB} illustrates this situation.\cite{RalfProc} 
The theory starts from perfect symmetry and the vacuum is true null space 
(Fig.~\ref{fig:SLSB}a).  
After the SSB, the vacuum is saturated by a scalar Higgs field $\ph$,  
and the particles obtain mass terms (Fig.~\ref{fig:SLSB}b). 
This idea can be extended to a vector field (Fig.~\ref{fig:SLSB}c). 
The ultra-high-energy theories, such as Planck-scale physics theories, 
have many Lorentz vector fields (or, more generally, Lorentz tensor fields). 
When the universe cools, 
if any of them acquire nonzero vacuum expectation values, 
then the vacuum can be saturated with vector fields.  
(Fig.~\ref{fig:SLSB}d). 
Note that, theoretically, 
SLSB is conceived to occur earlier than the SSB of the SM, 
unlike this cartoon. 
Such vector fields are the background fields of the universe, 
and they are fixed in space; 
couplings with the SM fields generate interaction terms in the vacuum. 

\beq
\L=i\ps\ga_\mu\partial^\mu\bar\ps
+m\ps\bar\ps
+\ps\ga_\mu a^\mu\bar\ps
+\ps\ga_\mu c^{\mu\nu}\partial_\nu\bar\ps+\cdots~.\no 
\eeq

In this expression, the coefficients $a^\mu$ and $c^{\mu\nu}$ represent 
vacuum expectation values of vector and tensor fields, 
and they correspond to background fields that fill the universe. 
The crucial observation is that, 
since these Lorentz tensors are fixed in space and time, 
they cause \textit{direction-dependent physics}. 
In particular, rotation of the Earth (period $\tsid$~sec) 
causes sidereal time dependent physics for any terrestrial measurement 
if the SM fields couple with Lorentz-violating background fields.
Therefore, the smoking gun of Lorentz violation is to find 
a sidereal time dependence in any physics observable.
 
\section{What is Lorentz and CPT violation?}

We introduce Lorentz violation as coupling terms between 
ordinary SM fields and background fields in the universe. 
They are Lorentz scalars of the coordinate transformation; 
however since background fields are fixed in space, 
motion of the SM particles generates coordinate-dependent physics.

The situation is illustrated in Figure~\ref{fig:LT}. 
The top cartoon (Fig.~\ref{fig:LT}a) shows our setting: 
a SM particle is moving in two-dimensional space, from bottom to top, 
as seen by the local observer (Einstein). 
The space is filled with a hypothetical Lorentz-violating background field, $a^\mu$ 
(depicted by arrows). 
There are two ways to move this particle from left to right for the local observer: 
Particle Lorentz transformation and Observer Lorentz transformation.

\subsection{Particle Lorentz transformation}

The first one is Particle Lorentz transformation, 
where the motion of a SM particle is actively transformed 
in the fixed coordinate system (Fig.~\ref{fig:LT}b). 
Since the background field is unchanged, as a consequence, 
a coupling between the SM particle and the background field is not preserved; 
therefore, one can see Lorentz violation. 
In other words, Lorentz violation generally means the Particle Lorentz violation, 
and it implies direction-dependent physics of SM particles in the fixed coordinate space.

\subsection{Observer Lorentz transformation}

\begin{figure}[tbp]
\centerline{\psfig{file=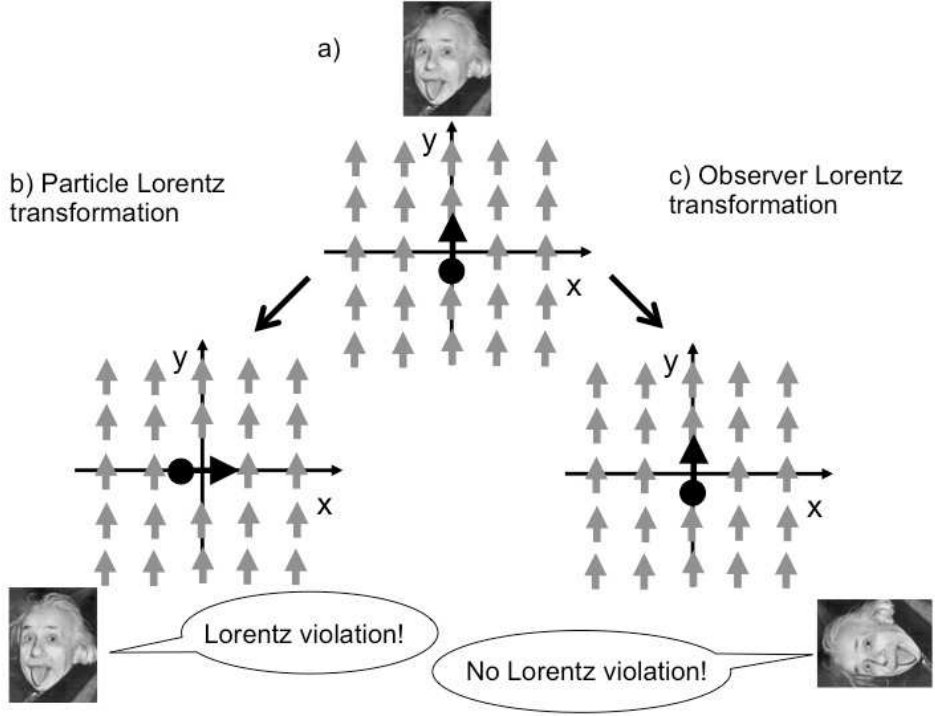,width=5.0in}}
\vspace*{8pt}
\caption{An illustration of Particle Lorentz transformation and 
Observer Lorentz transformation.}
\label{fig:LT}
\end{figure}

The second way is Observer Lorentz transformation, 
where the coordinate is inversely transformed 
(Fig.~\ref{fig:LT}c). 
In this cartoon, 
if Einstein (the local observer) turns his neck $90^\circ$ counterclockwise, 
the SM particle moves from left to right for the local observer. 
However, the background field is also transformed to the new coordinate system leaving 
the coupling with the SM particle unchanged.  
One cannot generate Lorentz violation by Observer Lorentz transformation, 
because Observer Lorentz transformation only corresponds to a coordinate transformation. 
In other words, coordinate transformations preserve the Lorentz-violating effect, 
and every observers agree with the same Lorentz-violating effect.

\subsection{CPT violation~\label{subsec:CPT}}

There is a close link between Lorentz symmetry and CPT symmetry. 
Here, ``CPT'' represents the combination of charge transformation (C), 
parity transformation (P), and time reversal (T). 
It is known that none of these, taken individually, is a symmetry of the SM, 
but from the CPT theorem\cite{CPT}  
we expect that their combination is a perfect symmetry. 
Since the CPT theorem is based on Lorentz symmetry, 
one can expect CPT violation 
when Lorentz invariance is broken. 
This is manifest in the appearance of CPT-odd terms in the Lagrangian, 
which appear as a subset of the terms that break Lorentz invariance. 
The phase of CPT transformation 
is related to the number of Lorentz indices, $i$, 
transforming under the Particle Lorentz transformation. 

\beq
CPT~phase = (-1)^i~.\no
\eeq

In the SM any Lagrangian is the linear sum of CPT-even, or phase$=+1$ terms. 
This is the main consequence of the CPT theorem, 
and this is why any Lagrangian is CPT invariant. 
However, if the theory includes Lorentz violation, 
it is possible that $i$ can be odd. 
When $i$ is odd number, Lorentz violation causes CPT violation, 
and this is called ``CPT-odd'' Lorentz violation. 
On the other hand, if $i$ is an even number, the phase of CPT transformation is even, 
and the theory does not violate CPT, even though it contains Lorentz violation. 
Such Lorentz violation is called ``CPT-even'' Lorentz violation. 
CPT-odd and CPT-even Lorentz violation 
differ only in the number of their Lorentz indices.

\begin{itemize}
\item coefficients of CPT-odd Lorentz-violating terms ($a^\mu$, $g^{\mu\nu\la}$, $\cdots$) 
\item coefficients of CPT-even Lorentz-violating terms ($c^{\mu\nu}$, $\ka^{\mu\nu\la\si}$, $\cdots$)
\end{itemize}

Note that the interactive quantum field theory necessarily violates Lorentz symmetry 
if CPT symmetry is not preserved.\cite{Greenberg} 
This general theory is consistent with the argument here 
and Standard-Model Extension (SME), 
which we discuss in the next section. 

\section{Analysis of Lorentz violation~\label{sec:analysis}}

Lorentz violation is realized as a coupling of SM particles and background fields. 
To specify the components of the Lorentz-violating vector or tensor fields, 
the coordinate system must be specified. 
Then the general Lagrangian, including all possible Lorentz-violating terms, is prepared. 
Finally, using this Lagrangian, 
observable physical quantities can be identified.

\subsection{The Sun-centered coordinate system~\label{subsec:coords}}

\begin{figure}[tbp]
\centerline{\psfig{file=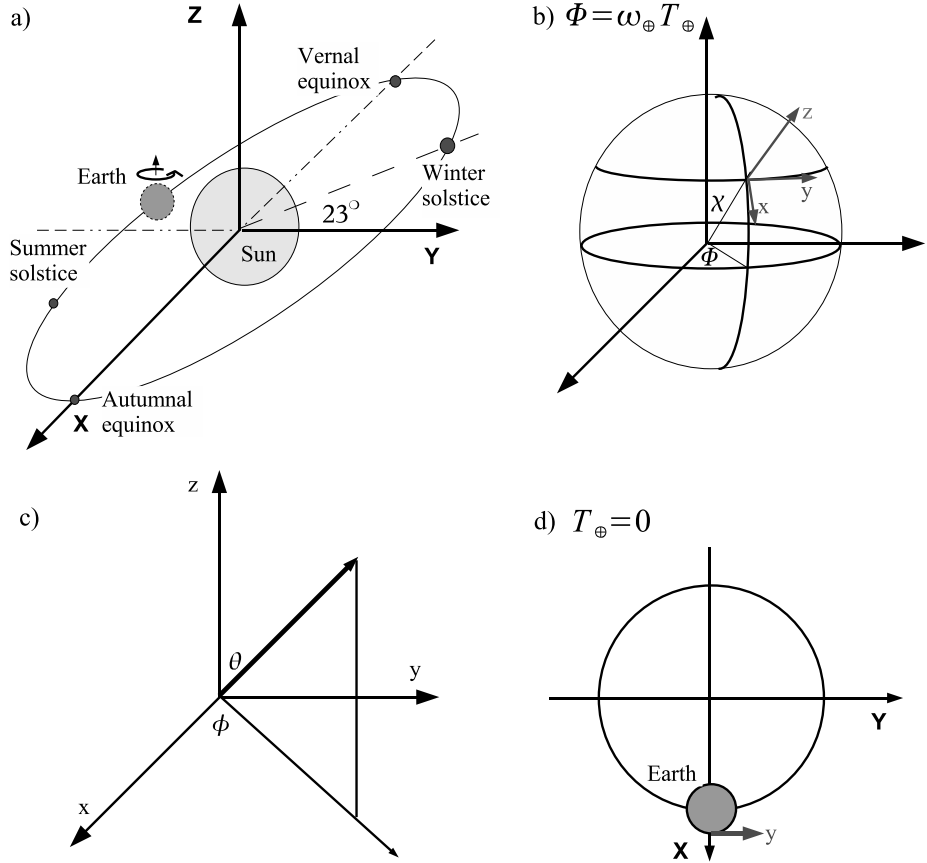,width=5.0in}}
\vspace*{8pt}
\caption{The coordinate system used by this analysis; 
(a) first, the motion of the Earth is described in Sun-centered coordinates, 
(b) then the local coordinates of the experiment site are described in Earth-centered coordinates, 
(c) finally, the direction of the neutrino beam is described in the local polar coordinate system. 
(d) The time zero is defined when the experiment site is at midnight near the autumnal equinox, 
in other words, when the large ``Y'' and small ``y'' axes almost align.}
\label{fig:coords}
\end{figure}

The choice of the coordinate system is arbitrary, 
since Particle Lorentz violation in one coordinate system is 
preserved in another coordinate system through Observer 
Lorentz transformation. 
Nevertheless, in order to compare experimental results from different experiments 
in a physically meaningful way, a common frame should be used. 
For this purpose, we need a universal coordinate system (Fig.~\ref{fig:coords}).\cite{LSND_LV}
The universal coordinate system is required to 
be reasonably inertial in our timescale. 
The Sun-centered coordinated system is just such a coordinate system (Fig.~\ref{fig:coords}a).
Here, the rotation axis of the Earth aligns with the orbital axis by 
tilting the orbital plane by $23.4^\circ$, defining the Z-axis. 
The X-axis points towards the vernal equinox, 
and the Y-axis completes the right-handed triad. 
Obviously, we assume Lorentz-violating field is uniform at least the scale 
of the solar system. 
This can be justified in many ways, for example, 
we know the weak and the electromagnetic laws 
are same in far stars through the observation, indicating Lorentz-violating 
fields are also uniform in these scales if they are arisen through the spontaneous breaking process.
Note, Sun-centered is more suitable than galaxy-centered, 
because although the galactic rotation is faster, 
the galactic rotation takes too long for human observation to change the direction,  
and consequently it cannot help test the violation of rotational symmetry. 
Then the location of the experiment is specified by 
the Earth-centered coordinate system (Fig.~\ref{fig:coords}b). 
Here, the x-axis points south, the y-axis points east, 
and the z-axis points to the sky from the site of the experiment. 
Finally, local polar coordinates specify the direction of the beam 
(Fig.~\ref{fig:coords}c). 
The time zero of the sidereal time is defined as being the position of the 
experiment at midnight near the autumnal equinox (Fig.~\ref{fig:coords}d). 

\subsection{Standard-Model Extension (SME)}

For the general search for Lorentz violation, 
the Standard-Model Extension (SME)\cite{SME1,SME2} is widely used by the community. 
Various types of data are analyzed under SME,\cite{SMEreview,SMEtable} 
including neutrino oscillation data.\cite{LSND_LV,MINOS_LV1,MINOS_LV2,MINOS_LV3,IceCube_LV} 
The SME is a self-consistent effective field theory including Particle Lorentz violation. 
In principle, SME is an infinite series of all types of interactions, 
but many analyses are limited to the renormalizable sector, called the minimal SME. 
For example, under the minimal SME, the effective Lagrangian for neutrinos can be written 
as,\cite{KM1}  

\beq
{\cal L} 
&=& 
\fr{1}{2}i{\bar {\ps}}_A{\Ga}^{\mu}_{AB}\stackrel{\leftrightarrow}
{D_{\mu}}{\ps}_{B}-{\bar {\ps}}_{A} M_{AB}{\ps}_{B}+h.c.,\\
\Ga_{AB}^{\nu}&\equiv&
\ganu\de_{AB}+c_{AB}^{\mu\nu}\gam+d_{AB}^{\mu\nu}\gaf\gam+e_{AB}^{\nu}+if_{AB}^{\nu}\gaf
+\fr{1}{2}g_{AB}^{\la\mu\nu}\si_{\la\mu},~\label{eq:gamma}\\
M_{AB}&\equiv&
m_{AB}+im_{5AB}\gaf+a_{AB}^{\mu}\gam+b_{AB}^{\mu}
+\fr{1}{2}H_{AB}^{\mu\nu}\si_{\mu\nu}.~\label{eq:mass}
\eeq

Here, the $AB$ subscripts represent Majorana basis flavor space ($6\times 6$ for convention). 
The first term of Eq.~\ref{eq:gamma} and the first and second terms of
Eq.~\ref{eq:mass} are the only nonzero terms in the SM, 
and the rest of the terms are from the SME. 
As we see, these SME coefficients can be 
classified into two groups (Sec.~\ref{subsec:CPT}), 
namely $e^{\mu}_{AB}$, $f^{\mu}_{AB}$,
$g^{\mu \nu \la}_{AB}$, $a^{\mu}_{AB}$, and $b^{\mu}_{AB}$ 
which are CPT-odd SME coefficients, 
and $c^{\mu \nu}_{AB}$, $d^{\mu \nu}_{AB}$, and $H^{\mu\nu}_{AB}$ 
which are CPT-even SME coefficients. 

\subsection{Lorentz-violating neutrino oscillations}

Once we have a suitable formalism, such as the SME, 
we are ready to write down physical observables. 
The effective Hamiltonian relevant for the $\nu-\nu$ oscillations can be written,~\cite{KM1}

\beq
(h_{\mathrm{eff}})_{ab} 
&\sim&
\fr{1}{|\vec{p}|}[(a_L)^\mu p_\mu-(c_L)^{\mu\nu}p_\mu p_\nu]_{ab}\label{eq:nu_hami}
\eeq

Here, $(a_L)_{ab}^{\mu}\equiv(a+b)_{ab}^{\mu}$ 
and $(c_L)_{ab}^{\mu\nu}\equiv(c+d)_{ab}^{\mu\nu}$. 
Note that we drop the neutrino mass term since the standard neutrino mass term 
is negligible for short baseline neutrino oscillation experiments, 
such as MiniBooNE. 

Solutions of this Hamiltonian have very rich physics, 
but for our purpose, we restrict ourselves to 
short-baseline $\numu-\nue$ ($\numubar-\nuebar$) oscillation phenomena. 
By assuming the baseline is short enough compared to the oscillation length, 
the $\numu-\nue$ oscillation probability can be written as follows,~\cite{KM3}

\beq
P_{\numu\to\nue} & \simeq & \fr{L^2}{(\hbar c)^2} |\, \C\indxn 
  +\As\indxn \sin\om_\oplus T_\oplus
  +\Ac\indxn \cos\om_\oplus T_\oplus \no\\
 & & 
  +\Bs\indxn \sin2\om_\oplus T_\oplus
  +\Bc\indxn \cos2\om_\oplus T_\oplus\,|^2.
\label{eq:nu_prob}
\eeq

Here, $\om_\oplus$ stands for the sidereal time angular frequency ($\om_\oplus=\frac{2\pi}{\tsid}$~rad/s), 
as opposed to the solar time angular frequency ($\om_\odot=\frac{2\pi}{\tgmt}$~rad/s). 
The neutrino oscillation is described by the function of the sidereal time $T_\oplus$, 
with the sidereal time independent amplitude $\C\indxn$, and 
the sidereal time dependent amplitudes, $\As\indxn$, $\Ac\indxn$, $\Bs\indxn$, and $\Bc\indxn$. 
Therefore, an analysis of Lorentz and CPT violation in neutrino oscillation data involves 
fitting the data with Eq.~\ref{eq:nu_prob} to find nonzero sidereal time dependent amplitudes.

In terms of the coefficients for Lorentz violation, 
these amplitudes are explicitly given by,

\beq
\C\indxn &=&\Ca\indxn+E\Cc\indxn  \label{eq:C}\no\\
\As\indxn&=&\Asa\indxn+E\Asc\indxn\label{eq:As}\no\\
\Ac\indxn&=&\Aca\indxn+E\Acc\indxn\label{eq:Ac}\\
\Bs\indxn &=& E\Bsc\indxn\no\\
\Bc\indxn &=& E\Bcc\indxn\no
\eeq

\beq
\Ca\indxn &=& (a_L)^T_{\indxn} + \nh^Z (a_L)^Z_{\indxn}
\label{eq:Ca}\no
\\
\Cc\indxn &=& -\frac{1}{2}(3-\nh^Z \nh^Z )(c_L)^{TT}_{\indxn} +2\nh^Z (c_L)^{TZ}_{\indxn}+\frac{1}{2}(1-3\nh^Z \nh^Z )(c_L)^{ZZ}_{\indxn}
\label{eq:Cc}\no
\\
\Asa\indxn &=& \nh^Y (a_L)^X_{\indxn}+\nh^X (a_L)^Y_{\indxn}
\label{eq:Asa}\no
\\
\Asc\indxn &=& -2\nh^Y (c_L)^{TX}_{\indxn} +2\nh^X (c_L)^{TY}_{\indxn} +2\nh^Y \nh^Z (c_L)^{XZ}_{\indxn}-2\nh^X \nh^Z (c_L)^{YZ}_{\indxn}
\label{eq:Asc}\no
\\
\Aca\indxn &=& -\nh^X (a_L)^X_{\indxn}+\nh^Y (a_L)^Y_{\indxn}
\label{eq:Aca}
\\
\Acc\indxn &=& 2\nh^X (c_L)^{TX}_{\indxn} +2\nh^Y (c_L)^{TY}_{\indxn} -2\nh^X \nh^Z (c_L)^{XZ}_{\indxn}-2\nh^Y \nh^Z (c_L)^{YZ}_{\indxn}
\label{eq:Acc}\no
\\
\Bsc\indxn &=& \nh^X \nh^Y ((c_L)^{XX}_{\indxn}-(c_L)^{YY}_{\indxn})-(\nh^X \nh^X -\nh^Y \nh^Y )(c_L)^{XY}_{\indxn}
\label{eq:Bsc}\no
\\
\Bcc\indxn &=& -\frac{1}{2}(\nh^X \nh^X -\nh^Y \nh^Y )((c_L)^{XX}_{\indxn}-(c_L)^{YY}_{\indxn}) -2\nh^X \nh^Y (c_L)^{XY}_{\indxn}
\label{eq:Bcc}\no
\eeq

Here, 
the $\nh^X$, $\nh^Y$, and $\nh^Z$ are the direction vectors of 
the neutrino beam in the Sun-centered coordinates (Sec.~\ref{subsec:coords}). 
The components are described with a co-latitude $\ch$  
of detector location in the Earth-centered system 
(Fig.~\ref{fig:coords}b), and the zenith and azimuthal angles 
$\th$ and $\ph$ of the local beam system. 
(Fig.~\ref{fig:coords}c):
\beq
\left(
\begin{array}{c}
\nh^X \\
\nh^Y \\
\nh^Z
\end{array}
\right)
&=&
\left(
\begin{array}{c}
\cos \ch \sin \th \cos \ph + \sin \ch \cos \th \\
\sin \th \sin \ph \\
-\sin \ch \sin \th \cos \ph + \cos \ch \cos \th 
\end{array}
\right)
\label{eq:dvec}
\eeq

For the antineutrino oscillation analysis, one needs to switch the sign of $a_L$
according to CPT-odd nature of CPT-odd coefficients ($a_L\to-a_L$).
 
In the reality of the analysis, fitting five parameters using Eq.~\ref{eq:nu_prob} is not easy. 
Therefore, we also consider the following three-parameter model, 
by setting $\Bs\indxn$ and $\Bc\indxn$ to be zero by hand. 
This model, Eq.~\ref{eq:nu_prob_3}, can be motivated, for example, by assuming nature only has CPT-odd 
SME coefficients.

\beq
P_{\numu\to\nue} & \simeq & \fr{L^2}{(\hbar c)^2}
|\C\indxn 
+\As\indxn \sin\om_\oplus T_\oplus
+\Ac\indxn \cos\om_\oplus T_\oplus|^2.
\label{eq:nu_prob_3}
\eeq

\subsection{Lorentz violation as an alternative neutrino oscillation model}

\begin{figure}[]
\centerline{\psfig{file=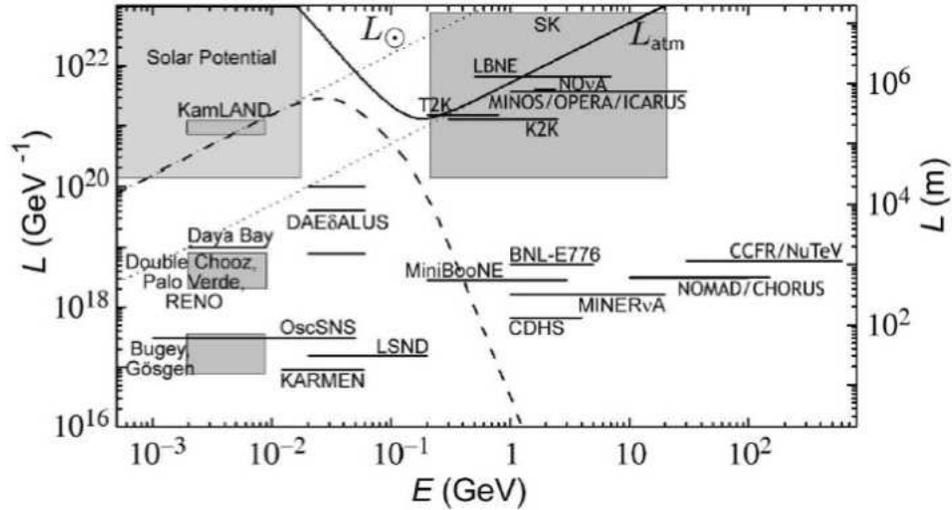,width=5.0in}}
\vspace*{8pt}
\caption{L-E diagram with the $\nu$SM (two straight dotted lines) 
and the Puma model (two dashed and solid curves).}
\label{fig:LEplot}
\end{figure}

Because of the unconventional energy dependence of Lorentz-violating terms in the Hamiltonian 
($E^0$ and $E^1$), 
naively, its energy dependence on neutrino oscillations is different from one expected 
from the three massive neutrino model (so-called $\nu$SM). 
However, it is also possible to ``mimic'' neutrino mass-like 
energy dependence ($E^{-1}$)
using Lorentz violating terms only.\cite{KM2,tandem,puma1,puma2} 
There is a chance that such types of models could be correct, 
because we currently have some tensions in the world neutrino oscillation data. 
For this purpose, it would be helpful to show the phase space of neutrino oscillations 
in a model-independent way. 
The L-E diagram (Fig.~\ref{fig:LEplot}) shows world's neutrino oscillation experiments mapped 
with their energy and baseline.\cite{puma1} 

The curves in Figure~\ref{fig:LEplot} represent the oscillation length. 
For example, massive neutrino oscillation solutions (=$L/E$ oscillatory dependence)  
are represented by the line $L\propto E$.
Here, data are consistent with two $L/E$ neutrino oscillations, 
the $\nuebar$ disappearance measurement at the KamLAND experiment (2 to 8 MeV), 
and the $\numu$ and $\numubar$ disappearance measurements at the 
long-baseline and atmospheric neutrino experiments (0.3 to 10 GeV). 
Therefore, we know 
there are \textit{at least two segments with ${\it L\propto E}$ on the L-E diagram}. 
Nevertheless, our knowledge outside of these segments is limited. 
There are proposed models, such as the Puma model,\cite{puma1,puma2}  
which have $L/E$ oscillatory dependencies in these energy ranges. 
So, here, the models are consistent with current data,\footnote{
Recent reactor neutrino disappearance oscillation results do not 
support Puma model~\cite{reactor1,reactor2,reactor3}} 
but outside of these energy ranges they have completely different dependencies. 
These alternative models are interesting because 
they have a chance to reproduce short-baseline anomalies, such 
as the MiniBooNE oscillation signals,\cite{MB_nu,MB_antinu} 
which we discuss in the next sections. 

\section{The MiniBooNE experiment}

\begin{figure}[tbp]
\centerline{\psfig{file=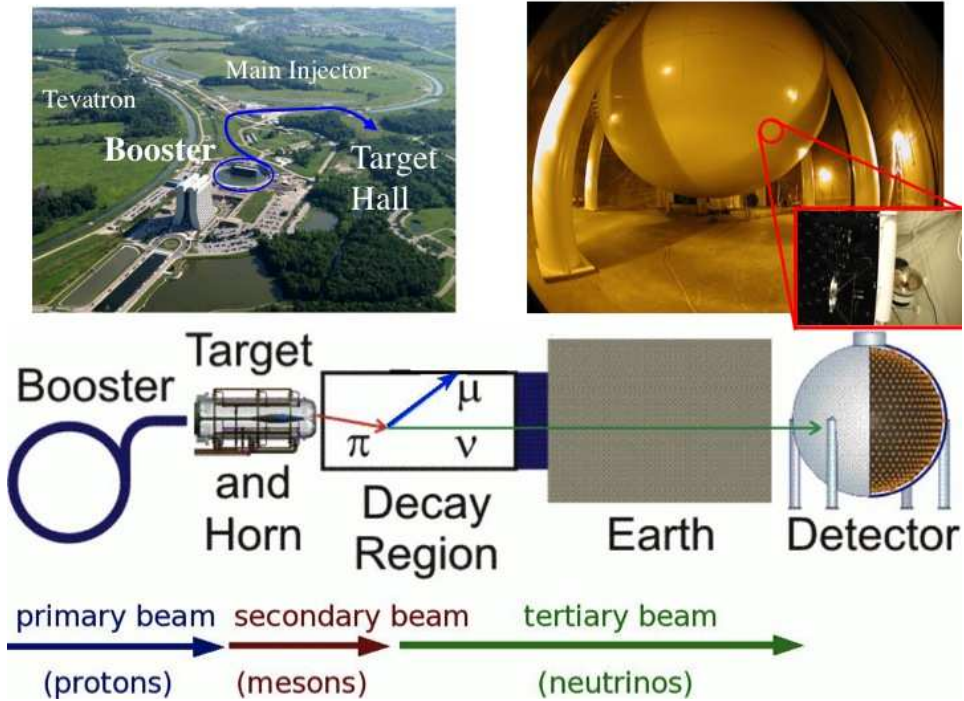,width=5.0in}}
\vspace*{8pt}
\caption{An overview of the MiniBooNE experiment. 
The top left picture shows the Fermilab site, 
including the Booster. 
The top right picture is the tank of the MiniBooNE detector. 
The bottom cartoon shows the sketch of the 
Booster Neutrino Beamline (BNB).}
\label{fig:MB}
\end{figure}

The MiniBooNE experiment is a short-baseline neutrino oscillation experiment 
at Fermilab, USA (2002-2012). Its primary goal is to find $\numu\to\nue$ ($\numubar\to\nuebar$) 
oscillations with an $\sim$800 (600)~MeV neutrino (antineutrino) beam with an $\sim$500~m baseline. 
Figure~\ref{fig:MB} shows the overview of the MiniBooNE experiment.\cite{Crimea} 

\subsection{Booster neutrino beamline (BNB)}

MiniBooNE uses neutrinos (antineutrinos) from the Booster neutrino beamline (BNB),\cite{MB_beam} 
which is illustrated in Figure~\ref{fig:MB}, bottom. 
The 8 GeV protons, the ``primary'' beam, are extracted from the Booster 
and steered to collide with the beryllium target in the magnetic focusing horn. 
The collision of protons and the target makes a shower of mesons, the ``secondary'' beam; 
and the toroidal field created by the horn focuses $\pi^+$ ($\pi^-$) 
for neutrino (antineutrino) mode. 
At the same time it defocuses $\pi^-$ ($\pi^+$), which create backgrounds. 
The decay-in-flight of $\pi^+$ ($\pi^-$) create $\numu$ ($\numubar$), 
the ``tertiary beam''.  
The consequent muon neutrinos (muon antineutrinos) are a wideband beams 
peaked in around 800 (600) MeV. 

\subsection{MiniBooNE detector}

The MiniBooNE detector is located 541~m to the north of the target.\cite{MB_detec} 
The detector is a 12.2~m diameter spherical Cherenkov detector, 
filled with mineral oil. 
An optically separated inner black region is covered with 
1,280 8-inch PMTs, and an outer white region has 240 8-inch PMTs which act as a veto 
(Fig.~\ref{fig:MB}, top right). 
The black inner cover helps to reduce reflections, 
so that one can reconstruct particles from the Cherenkov light more precisely; 
the outer white cover helps to enhance reflections,  
so that a smaller number of veto PMTs can cover a larger area. 

\subsection{Events in detector}

\begin{figure}[tbp]
\centerline{\psfig{file=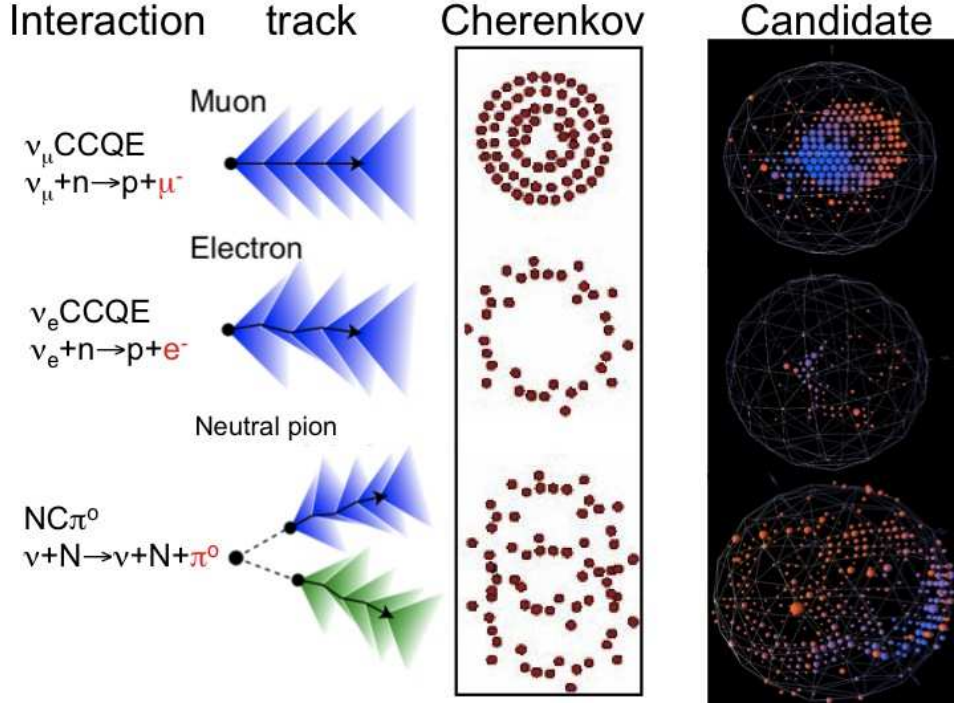,width=5.0in}}
\vspace*{8pt}
\caption{The particle type and characteristics. 
From left to right, interaction types, characteristics of tracks and Cherenkov rings, 
and event displays of candidate events.}
\label{fig:MB_topology}
\end{figure}

The time and charge information of the Cherenkov ring from the charged particle 
is used to estimate particle type, energy, and direction.\cite{MB_recon} 
For example, an electron-like track is characterized by a fuzzy-edged Cherenkov ring, 
compared with a sharp-edged muon-like Cherenkov ring. 
Based on particle type hypothesis, 
the track fitter estimates a particle energy and direction.  
Figure~\ref{fig:MB_topology} shows typical particles and their characteristic tracks, 
Cherenkov rings, and event candidates from the event display.\cite{Crimea} 

A variety of track fitters are developed to measure 
the kinematics of specific types of interactions. \cite{NCpi0_PRD,NCEL_PRD,CCpip_PRD,CCpi0_PRD} 
Among them, the most important reaction for the oscillation analysis is 
the charged current quasielastic (CCQE) interaction,\cite{MB_CCQEPRL,MB_CCQEPRD} 
which is characterized by one outgoing charged lepton 
(at the BNB energy, protons seldom exceed a Cherenkov threshold of $\sim$350~MeV kinetic energy). 
If a charged lepton is detected from the CCQE interaction, 
one can reconstruct the neutrino energy, $E_\nu^{QE}$, 
by assuming the target nucleon is at rest and 
the interaction type is truly CCQE\cite{MB_CCQEPRL} (QE assumption). 

\beq
E_\nu^{QE}
&=&
\fr{2(M_n-B)E_\mu-((M_n-B)^2+m_\mu^2-M_p^2)}
{2\cdot[(M_n-B)-E_\mu+\sqrt{E_\mu^2-m_\mu^2}\cos\th_\mu]}.\label{eq:recEnu}
\eeq

Here, $M_n$, $M_p$, and $m_\mu$ are the neutron, proton, and muon masses, 
$E_\mu$ is the total muon energy, 
$\th_\mu$ is the muon scattering angle, 
and $B$ is the binding energy of carbon. 
Ability to reconstruct neutrino energy is essential for neutrino oscillation physics, 
since neutrino oscillations are function of neutrino energy.

It is interesting to note that many of the neutrino interaction cross sections 
measured by MiniBooNE are at a higher rate and harder spectrum than historically known 
values and disagree with interaction models tuned with 
old bubble chamber data.\cite{Benhar}  
This fact triggered the development of a new class of 
neutrino interaction models,\cite{mec1,mec2,mec3,tem,rgf,sf,dwia} 
mostly by including nucleon correlations. 
These new models even question how to reconstruct neutrino energy\cite{mec_enu1,mec_enu2} 
with the QE assumption, traditionally done in all Cherenkov-type detectors. 
Therefore, similar to other fields (e.g., cosmology), 
the further study of neutrino physics just increases the number of mysteries!

\subsection{Oscillation analysis}

The signature of the $\numu\to\nue$ ($\numubar\to\nuebar$) oscillation is the 
single, isolated electron-like Cherenkov ring produced by the CCQE interaction. 
\beq
\numu&\stackrel{oscillation}{\longrightarrow}&\nue+n\to\mu^++p~,\no\\
\numubar&\stackrel{oscillation}{\longrightarrow}&\nuebar+p\to\mu^-+n~.\no
\eeq 

Note, since MiniBooNE was not magnetized, 
electrons and positrons were not distinguished. 
Thus, the analysis of neutrino mode and antineutrino mode is reasonably parallel, 
except for some differences in handling backgrounds. 

There are two backgrounds which contribute equally to our signals. 
The first class is the ``misID'', and this is dominated by 
a single gamma ray from the neutral current channels,  
such as radiative $\De$ decay and $\pi^\circ$ production, 
where only one gamma ray is detected. 
It is essential to constrain our misID background predictions on those channels 
using measurements of controlled samples. 
For this purpose, we measured the neutral current $\pi^\circ$ production rate 
{\it in situ}, and the result is used to tune $\pi^\circ$ kinematics in our simulation. 
We also used the measured $\pi^\circ$ production uncertainty in our simulation.\cite{NCpi0_PLB} 

Another major background is the intrinsic background, namely $\nue$ ($\nuebar$) as 
beam contamination. 
Although they can be predicted by the beamline simulation, 
and are expected to be $<$0.5\%, 
this is a critical background for the $\sim$0.5\% 
appearance oscillation search carried out by experiments, such as MiniBooNE.  
Again, {\it in situ} measurements largely help to reduce errors in the simulation. 
For example, the majority of $\nue$ ($\nuebar$) are from $\mu$-decay in the beamline but 
one can constrain their variations from the measured $\numu$ ($\numubar$) rate, 
where both $\nue$ and $\numu$ ($\nuebar$ and $\numubar$) 
are related through the $\pi^+$ ($\pi^-$) decay 
(for $\pi^+$, $\pi^+\to\numu\mu^+~,\mu^+\to\numubar\nue e^+$). 
Another major source of $\nue$ ($\nuebar$) is kaon decay. 
MiniBooNE utilizes SciBooNE experiment data to constrain it.\cite{SB_kaon} 
SciBooNE is a tracker for the neutrino cross section measurement, located upstream of MiniBooNE, 
and their precise track measurement is sensitive to primary mesons in the beamline. 
More specifically, K-decay origin neutrinos are higher energy, and tend to make multiple 
tracks in the SciBooNE detector. 
This information provides the constraint on the errors on predictions of 
$\nue$ ($\nuebar$) from K-decay in MiniBooNE.

After the evaluation of all backgrounds, 
MiniBooNE finds a signal-to-background ratio of roughly one to three, 
with expected oscillation parameters. 

\subsection{MiniBooNE neutrino mode oscillation result}

\begin{figure}[tbp]
\centerline{\psfig{file=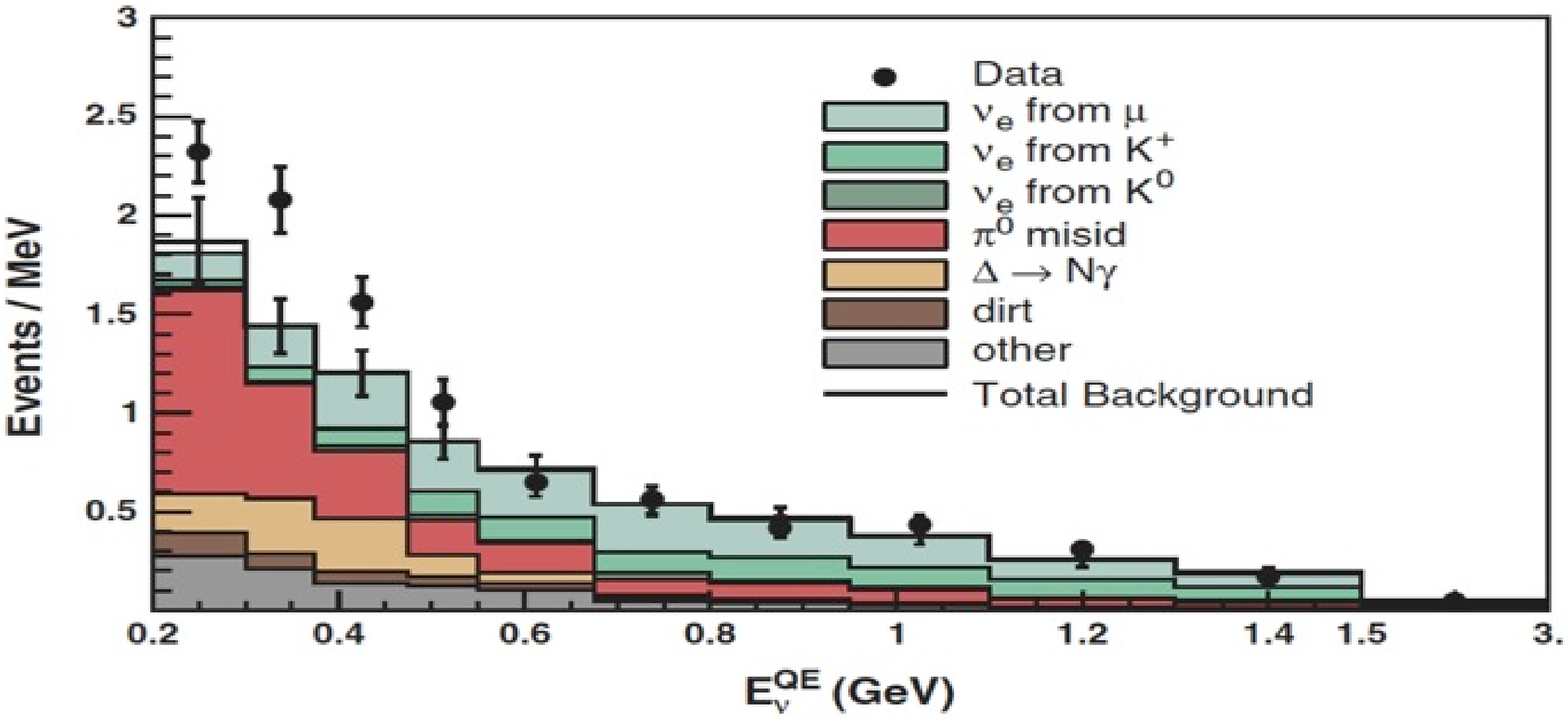,width=4.0in}}
\vspace*{8pt}
\caption{MiniBooNE neutrino mode $\nue$ appearance search result. 
The data points with errors are shown together with predicted backgrounds.}
\label{fig:MB_nu}
\end{figure}

For neutrino mode data analysis,\cite{MB_nu} we use $\np$  protons on target (POT) data. 
After all cuts, an excess of $\nue$ candidate events 
in the ``low-energy'' region (200$<E_\nu^{QE}$(MeV)$<$475) was observed (Fig.~\ref{fig:MB_nu}).  
A total of $\nln$ events are observed in this region, as compared to the predicted 
$\nlb\pm\nlt$(stat.)$\pm\nly$(syst.). 
Interestingly, this excess does not show the expected $L/E$ 
energy dependence of a simple two massive neutrino oscillation model. 
Therefore, this excess might be new physics.

\subsection{MiniBooNE antineutrino mode oscillation result}

\begin{figure}[tbp]
\centerline{\psfig{file=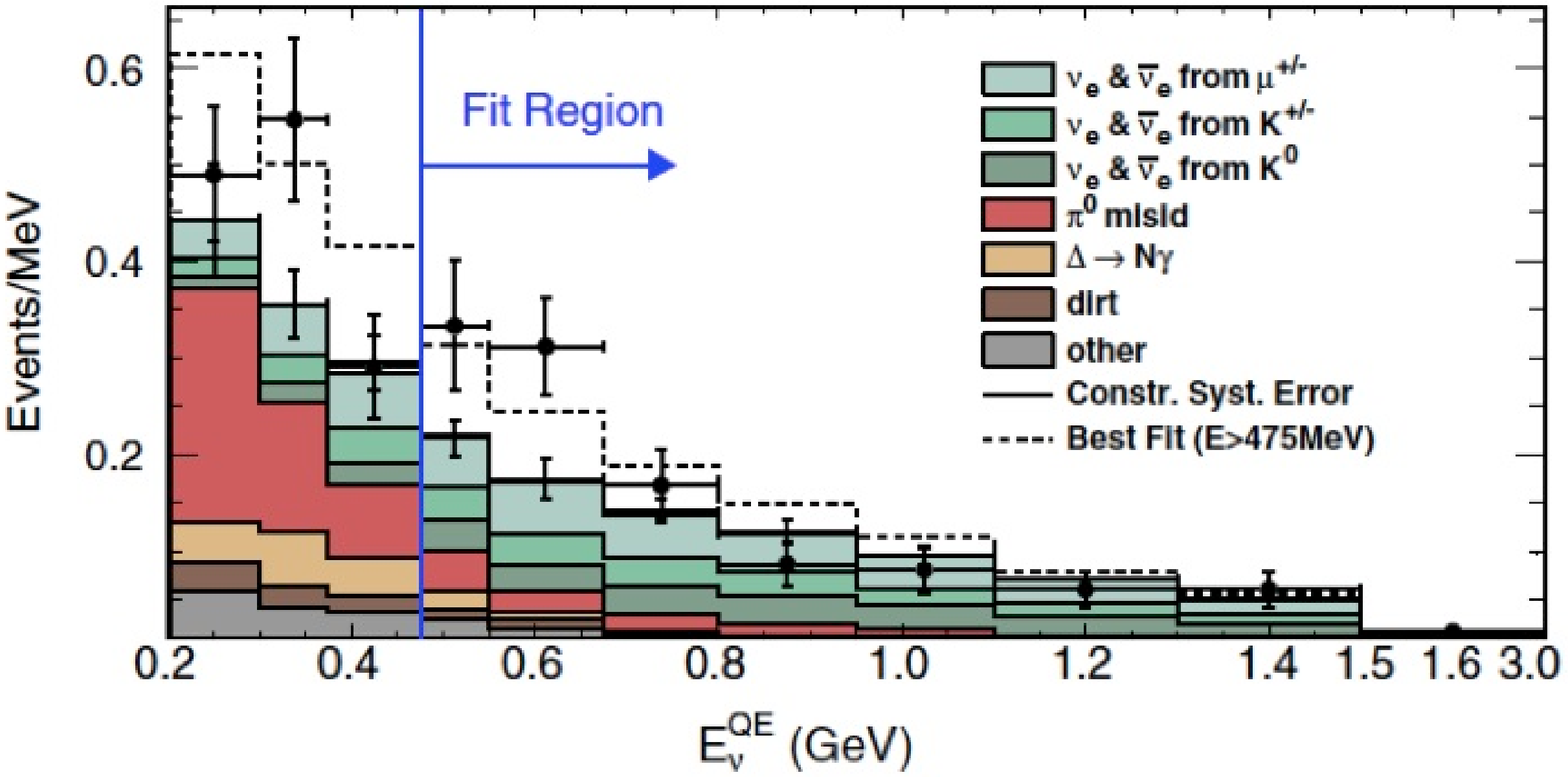,width=4.0in}}
\vspace*{8pt}
\caption{MiniBooNE antineutrino mode $\nuebar$ appearance search result.
The data points with errors are shown together with predicted backgrounds.}
\label{fig:MB_antinu}
\end{figure}

For the antineutrino mode analysis,\cite{MB_antinu} we use $\ap$~POT data. 
Here, MiniBooNE not only observed an excess in the low energy region, 
an excess in the ``high-energy'' region (475$<E_\nu^{QE}$(MeV)$<$1300) was also observed 
(Fig.~\ref{fig:MB_antinu}).  
Therefore, in the ``combined'' region (200$< E_\nu^{QE}$(MeV)$<$1300), 
MiniBooNE observed $\atn$ $\nuebar$ candidate events as compared to the predicted 
$\atb\pm\att$(stat.)$\pm\aty$(syst.). 
Again, the $\nu$SM does not predict this excess, 
which therefore has the potential to be new physics.

\section{Lorentz-violating neutrino oscillation analysis in MiniBooNE}

In this section we follow the procedure described in Sec.~\ref{sec:analysis}.

\subsection{The coordinate system}

First, we make the time distribution of neutrino events from a standard GPS time stamp. 
The analysis is based on the sidereal time distribution, 
but we also use the local solar time to check time-dependent systematics. 
The coordinate system is described in Fig.~\ref{fig:coords}. 
The local coordinates of the BNB are specified by three angles,  
the co-latitude $\ch=\MBch^\circ$, 
the polar angle $\th=\MBth^\circ$, 
and the azimuthal angle $\ph=\MBph^\circ$. 

\begin{figure}[tbp]
\centerline{\psfig{file=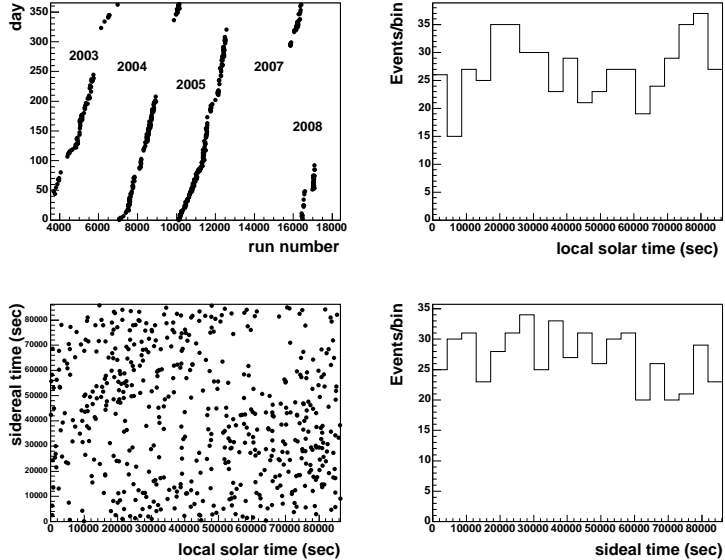,width=4.0in}}
\vspace*{8pt}
\caption{Time distribution of the MiniBooNE low energy excess in neutrino mode. 
The top left figure shows the day and run number of each event. 
The bottom left plot is the scatter plot of each event with sidereal and local solar time; 
projections of the distribution onto each axes are show to the right.}
\label{fig:nu_time}
\end{figure}

\begin{figure}[tbp]
\centerline{\psfig{file=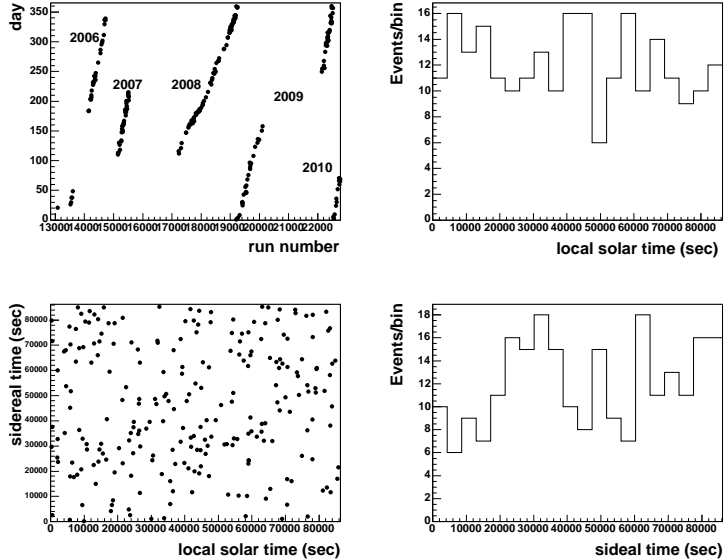,width=4.0in}}
\vspace*{8pt}
\caption{Time distribution of the MiniBooNE excess in antineutrino mode. 
The keys are same with Figure~\ref{fig:nu_time}.}
\label{fig:antinu_time}
\end{figure}

\subsection{Time dependent systematics}
 
\begin{figure}[tbp]
\centerline{\psfig{file=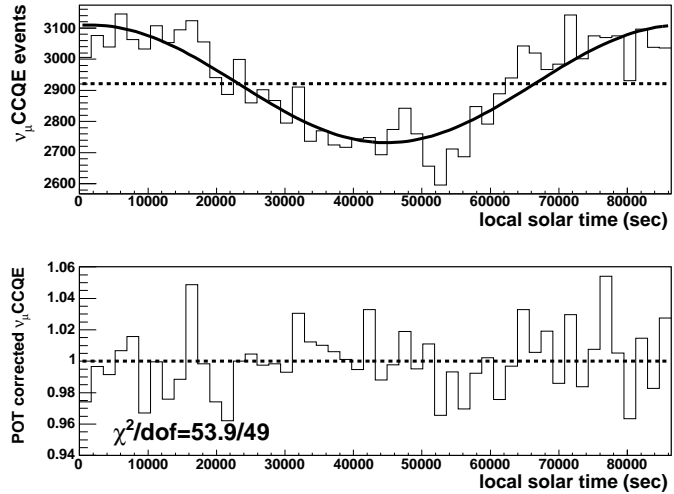,width=4.0in}}
\vspace*{8pt}
\caption{
The top histogram shows the $\numu$CCQE local solar time distribution, 
with the arbitrary normalized fit curve extracted from the POT local solar time distribution 
in the same time period (solid curve). 
The bottom plot shows the same events after correcting for variations in POT. 
The $\ch^2$ of this distribution with a flat hypothesis is $\tdch/\tddf$ 
($\tdpb$\% compatibility).}
\label{fig:tdist}
\end{figure}

The signature of the Lorentz violation in this analysis is 
the sidereal time-dependent neutrino oscillations. 
However, any solar time distribution may show up in the sidereal time distribution, 
if data taking is not completely continuous. 
This is the case for MiniBooNE, as you see from top left plots of 
Figure~\ref{fig:nu_time} and ~\ref{fig:antinu_time}.  
There were occasional shutdowns of the accelerator, 
and data taking was not continuous over any of the years. 
Therefore, we need to check time-dependent systematics, 
which include any day-night effects of the detector and the beam. 
For example, electronics may have a higher efficiency in cold nighttime than hot daytime, etc. 
The quick way to evaluate all of these effects together is to utilize the neutrino data itself. 
The high-statistics $\numu$CCQE\cite{MB_CCQEPRD} data are used 
to evaluate day-night variations of the neutrino interaction rate. 
Figure~\ref{fig:tdist}, above, 
shows the variation of $\numu$CCQE events in local solar time.\cite{MB_LV} 
As you see, the $\numu$CCQE sample shows day-night variations. 
These originate from the variation of the beam. 
The number of POT shows sinusoidal curve, namely maximum POT is  
at midnight and decreases towards the daytime: 
presumably the beamline is occasionally accessed in the daytime, 
and cumulative human behavior over several 
years make a nice sinusoidal curve in POT, and in neutrino rate!
After correcting POT, 
$\numu$CCQE event shows flat (Fig.~\ref{fig:tdist}, bottom). 

Then, the question is how much the solar time variation of the POT 
actually shows up in the oscillation candidate sidereal variation. 
We found that a small variation persists in the $\numu$CCQE sidereal distribution. 
We evaluated the impact of this on our analysis by correcting 
the POT variation event by event in $\nue$ candidate data. 
It turned out that, because of the low statistics of oscillation candidate events and the smearing effect 
from solar time distribution to sidereal time distribution, 
the correction only had a negligible effect.  
Thus, we decided to use uncorrected events in later analysis. 

Similar study is needed for the antineutrino mode, 
but the conclusion is likely to be the same due to lower statistics 
of $\nuebar$ sample.

\subsection{Unbinned likelihood fit} 

To find best-fit (BF) parameters to describe MiniBooNE $\nue$ ($\nuebar$) candidate data 
with Lorentz violation, we employed an unbinned likelihood fit. 
The likelihood function $\La$ has following expression with two probability density functions (PDFs).

\beq
\Lambda &=& {{e^{-(\mu_s+\mu_b)}} \over {N!}} \prod_{i=1}^{N} 
(\mu_s \mF_s^i + \mu_b \mF_b^i) 
\times {1 \over {\sqrt{2\pi {\si_b}^2}}}\exp 
\left( -{{(\bar\mu_b-\mu_b)}^2 \over {2 {\si_b}^2}} \right) 
\label{eq:gulf}
\eeq
\begin{enumerate}
\item[N,] the number of observed candidate events 
\item[$\mu_s$,] the predicted number of signal events, the function of fitting parameters 
\item[$\mu_b$,] the predicted number of background events, floating within $1\si$ range
\item[$\mF_s$,] the PDF for the signal, the function of sidereal time and fitting parameters
\item[$\mF_b$,] the PDF for the background, not the function of the sidereal time
\item[$\si_b$,] the $1\si$ error on the predicted background 
\item[$\bar\mu_b$,] the central value of the predicted total background events
\end{enumerate}

This method is suitable because it has the highest statistical power for a low-statistics sample. 
The computation is performed to maximize this function. 
But in the reality, we maximize the log likelihood function.  
The maximum log likelihood (MLL) point provide the best fit (BF) parameter set. 
Then the constant surface of log likelihood function provide the errors.  
Neither the neutrino nor the antineutrino mode data allow us to extract errors 
if we fit all five parameters at once (Eq.~\ref{eq:nu_prob}), 
due to the  high correlation of parameters. 
Therefore, we focus on three-parameter fit (Eq.~\ref{eq:nu_prob_3}) to discuss errors and limits. 
Since the five-parameter fit is quantitatively similar to the three-parameter fit, 
we will focus the discussion of these results on the three-parameter fit.

\section{Results} 

This section describes the results of the fits.

\subsection{Fit result of neutrino mode data} 

\begin{figure}[tbp]
\centerline{\psfig{file=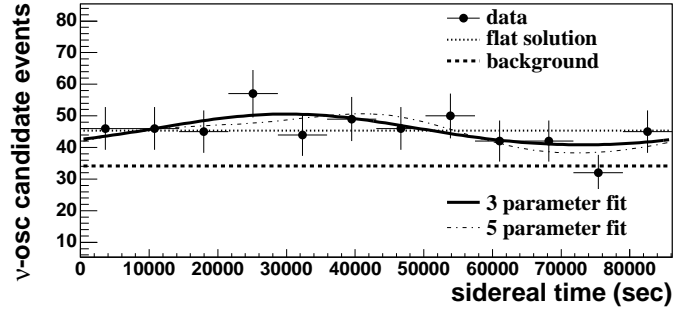,width=4.0in}}
\vspace*{8pt}
\caption{
The fit results for the neutrino mode low energy region. 
The plot shows the curves corresponding to the 
flat solution (dotted),  
three-parameter fit (solid),   
and five-parameter fit (dash-dotted), 
together with binned data (solid marker). 
Here the fitted background is shown as a dashed line, 
however the best fit for the background is 1.00 
(the default prediction).
}
\label{fig:nu_3F_lowE}
\end{figure}

Figure~\ref{fig:nu_3F_lowE} shows the neutrino mode low energy region fit results.\cite{MB_LV} 
The solid and dash-dotted lines are the best fit curves from three- and five-parameter fit. 
The dotted line is the flat solution. 
Since the fit is dominated by the $\C\indxn$ parameter 
(sidereal time-independent amplitude), 
both three- and five-parameter fit solutions have 
small time-dependent amplitudes and look like the flat solution.  
The details of the fit results were tested by using a fake data set. 
We constructed a fake data set with signal to 
evaluate errors of fit parameters of three-parameter fit 
($\C\indxn$, $\As\indxn$, and $\Ac\indxn$).  
Then we constructed a fake data set without signal, 
to evaluate the compatibility with flat solution 
over three parameter fit solution by $\De\ch^2$ method. 
It turns out data is compatible with flat solution over a $\Dcsqnu$\%, 
and it concludes $\nue$ candidate data is consistent with flat.  

\subsection{Fit result of antineutrino mode data} 

\begin{figure}[tbp]
\centerline{\psfig{file=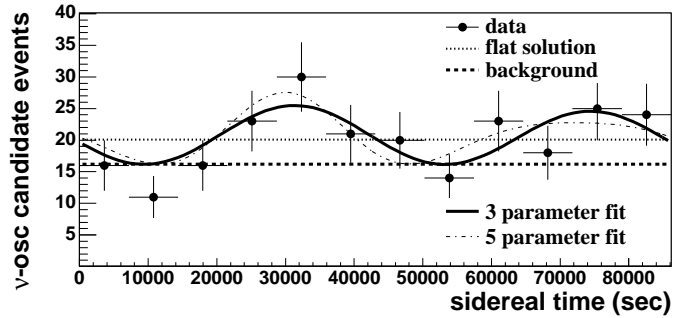,width=4.0in}}
\vspace*{8pt}
\caption{
The fit results for the antineutrino mode combined region. 
Notations are the same as Fig.~\ref{fig:nu_3F_lowE}. 
The fitted background is shown as a dashed line, 
and the best fit for the background is 0.97 
(3\% lower than the default prediction).
}
\label{fig:nub_3F_totE}
\end{figure}

Figure~\ref{fig:nub_3F_totE} shows the fit result 
for combined energy region in antineutrino mode, 
which is analogous with Figure~\ref{fig:nu_3F_lowE}.\cite{MB_LV} 
For antineutrino mode, the combined region is more interesting due to lower statistics. 
The fit result is more curious in antineutrino mode,  
because now the $\C\indxn$ parameter no longer significantly deviates from zero, 
but the fit favors a nonzero solution for 
the $\As\indxn$ and $\Ac\indxn$ parameters. 
The fit solutions look more different from the flat distribution. 
We again constructed a fake data set to find the significance of this solution, 
and it turns out that the compatibility with the flat solution is now only $\Dcsqnubar$\%. 
Although this is interesting, however the significance is not high enough to claim 
discovery.  

\subsection{Summary of the fit} 

In this analysis, three parameters, 
$\C\indxn$, $\As\indxn$, and $\Ac\indxn$ for neutrino and antineutrino mode are obtained by fits. 
Fits provide the BF values of above three parameters, 
as well as 1$\si$ errors and 2$\si$ limits.  
The expressions of these parameters are found in Eqs.~\ref{eq:Ac},~\ref{eq:Aca}, and~\ref{eq:dvec}. 
From the 2$\si$ limits we obtain 
we estimate the limit of each SME coefficient 
by setting all but one of SME coefficient as nonzero. 
Table~\ref{tab:smallSME} is the result. 
As you see, the limits of the SME coefficients from 
the MiniBooNE data are of the order of $10^{-20}$~GeV (CPT-odd), 
and $10^{-20}$ to $10^{-19}$ (CPT-even). 
Similar analysis have been done for the LSND data.\cite{CPT10_TK} 
However, these limits exclude 
any SME coefficients needed to explain the LSND data. 
Therefore, a simple picture using Lorentz violation 
to explain both LSND and MiniBooNE leaves some tension, 
and a mechanism to cancel the Lorentz-violating effect 
at high energy\cite{KM2,tandem,puma1,puma2} is needed.

\begin{table}
\tbl{List of SME coefficient limits,  
derived from 2$\si$ limits of fitting parameters, setting all but one of the SME coefficients to be zero.}
{\begin{tabular}{@{}lcclc@{}} \toprule
\hline
Coefficient&e$\mu$ ($\nu$ mode low energy region)&e$\mu$ ($\nubar$ mode combined region)\\
\hline
Re$\aL^T   $ or Im$\aL^T   $&4.2$\times 10^{-20}$~GeV&2.6$\times 10^{-20}$~GeV\\
Re$\aL^X   $ or Im$\aL^X   $&6.0$\times 10^{-20}$~GeV&5.6$\times 10^{-20}$~GeV\\
Re$\aL^Y   $ or Im$\aL^Y   $&5.0$\times 10^{-20}$~GeV&5.9$\times 10^{-20}$~GeV\\
Re$\aL^Z   $ or Im$\aL^Z   $&5.6$\times 10^{-20}$~GeV&3.5$\times 10^{-20}$~GeV\\
Re$\cL^{XY}$ or Im$\cL^{XY}$&---                     &---                     \\
Re$\cL^{XZ}$ or Im$\cL^{XZ}$&1.1$\times 10^{-19}$    &6.2$\times 10^{-20}$    \\
Re$\cL^{YZ}$ or Im$\cL^{YZ}$&9.2$\times 10^{-20}$    &6.5$\times 10^{-20}$    \\
Re$\cL^{XX}$ or Im$\cL^{XX}$&---                     &---                     \\
Re$\cL^{YY}$ or Im$\cL^{YY}$&---                     &---                     \\
Re$\cL^{ZZ}$ or Im$\cL^{ZZ}$&3.4$\times 10^{-19}$    &1.3$\times 10^{-19}$    \\
Re$\cL^{TT}$ or Im$\cL^{TT}$&9.6$\times 10^{-20}$    &3.6$\times 10^{-20}$    \\
Re$\cL^{TX}$ or Im$\cL^{TX}$&8.4$\times 10^{-20}$    &4.6$\times 10^{-20}$    \\
Re$\cL^{TY}$ or Im$\cL^{TY}$&6.9$\times 10^{-20}$    &4.9$\times 10^{-20}$    \\
Re$\cL^{TZ}$ or Im$\cL^{TZ}$&7.8$\times 10^{-20}$    &2.9$\times 10^{-20}$    \\
\hline
\end{tabular}\label{ta1}}
\label{tab:smallSME}
\end{table}

\section*{Conclusion}

Lorentz and CPT violation is a predicted signal at the Planck scale, 
and there are worldwide efforts to search for it. 
Neutrino oscillation is a natural interferometer, and the sensitivity to 
Lorentz violation is comparable with precise optical experiments. 
The MiniBooNE short-baseline neutrino oscillation experiment at Fermilab observed 
a $\nue$ ($\nuebar$) candidate event excess from a $\numu$ ($\numubar$) beam. 
These excesses are not understood by $\nu$SM; 
therefore, they might be the first signal of new physics. 
Since Lorentz-violation-motivated phenomenological neutrino oscillation models, 
such as the Puma model, predict an oscillation signal for the MiniBooNE, 
it is interesting to check the sidereal variation of MiniBooNE oscillation candidate data. 
The analysis found that neutrino mode data prefer a sidereal time independent solution, 
and the data is compatible with a flat distribution over  $\Dcsqnu$\%. 
The antineutrino mode prefers a sidereal time dependent solution, 
and the data is compatible with a flat solution only $\Dcsqnubar$\%, 
making this solution very interesting; 
however, the statistical significance is not high enough to claim as evidence. 
Since the data set we used for this analysis is about 
$\sim$50\% of the total antineutrino mode data set, 
reanalysis including full data set may increase the significance. 
Finally, from the fits, we extract limits of each minimal SME coefficient. 
The results from MiniBooNE leave tension with LSND 
under the simple Lorentz violation motivated model.  

\section*{Acknowledgments}

I thank Jorge D\'{i}az for valuable comments, 
I also thank Janet Conrad, Josh Spitz, Ben Jones, 
and Clemmie Jones for their careful reading of this manuscript. 
This work is supported by NSF PHYS-084784. 


\end{document}